\documentclass[epj]{svjour}
\usepackage{graphics}
\usepackage{amsmath}
\usepackage{braket}
\usepackage[toc,page]{appendix}
\usepackage{mathtools}
\usepackage{extarrows}

\usepackage{ucs}
\usepackage[utf8]{inputenc}
\DeclareMathOperator\arctanh{arctanh}

\newcommand{\be}{\begin{equation}}
\newcommand{\ee}{\end{equation}}
\newcommand{\bea}{\begin{eqnarray}}
\newcommand{\eea}{\end{eqnarray}}

\begin{document}
\title{Decaying universes and the emergence of Bell-type interuniversal entanglement in varying fundamental constants cosmological model}

\titlerunning{Decaying universes and the emergence of Bell-type interuniversal entanglement}


\author{Adam Balcerzak\inst{1,2} \and Mateusz Lisaj\inst{3}
}

\authorrunning{A. Balcerzak, M. Lisaj}
%
%
\institute{Institute of Physics, University of Szczecin,
Wielkopolska 15, 70-451 Szczecin,  Poland \and Copernicus Center for Interdisciplinary Studies, Szczepa\'nska 1/5, 31-011 Krak\'ow, Poland \and  Chair of Physics and Chemistry, Maritime University of Szczecin, Wa{\l }y Chrobrego 1-2, 70-500 Szczecin, Poland}
\date{Received: date / Revised version: date}
%
\abstract{In this paper, we consider a high-curvature limit of the varying fundamental constants toy model in which both the value of the speed of light and the value of the gravitational constant are related to the values of the two non-minimally coupled scalar fields. The high-curvature limit motivates the application of the third quantization procedure to such a toy model which results in a theory that describes bosonic massive particles that move freely in the three-dimensional minisuperspace associated with the degrees of freedom of the original model. Motivated by the idea that in the quantum cosmological description the minisuperspace gets promoted to a real configurational space of the system we supplement the third quantized action of the considered model with an interaction term that allows for decay and scattering processes. We show that such interaction term induces a scenario in which a parent universe decays into two universes described by a nearly maximally entangled Bell state. We eventually asses the strength of the entanglement, in the created pair of universes, by calculating the von Neumann entropy of entanglement.
} 
\maketitle

\section{Introduction}
\label{intro}

The idea of the multiverse is quite capacious. Most of the conceptualizations of such an idea can be assigned to one of Tegmark's levels of the multiverse \cite{tegmark}. Level one defines multiverse as regions beyond our cosmic horizon. The effects of quantum entanglement between different causally disconnected patches of space-time were investigated in \cite{Holman1,Holman2}. Tegmark's level two defines multiverse as a collection of post-inflationary bubbles with possibly different values of the physical constants. Level three involves Everett’s many-worlds interpretation of quantum physics while level four also includes all well defined mathematical structures. An interesting class of models, that can be classified as a level two multiverse, is based on the third quantization of the Wheeler-DeWitt wave function. In the process of the third quantization, the Wheeler-DeWitt wave function becomes an operator, which creates or annihilates universes characterized by some sets of quantum numbers (these are usually the momentum components in the minisuperspace) \cite{Robles_ent1,Robles_ent2}.
Such approach, which explicitly uses the third quantization formalism to describe the emergence of the entanglement in the multiverse, has inspired the area of research that explored different scenarios of generation of the interuniversal entanglement. It exploits the two different representations (the emergence of the entanglement resulting from the representation change is a generic feature of any quantum field theory \cite{Mukhanov}), namely the invariant one (for the first time found in \cite{Lewis}) which conserves the number of the universes and the diagonal one.  
Further development of the multiverse models relying on the third quantization includes postulating an interaction between the different universes which constitute the multiverse with each universe being represented by a different Wheeler-DeWitt wave function. It is usually assumed that the interaction between the universes can be represented by a term that formally describes an interaction between two harmonic oscillators coupled by a spring \cite{Serrano,Robles1,Kraemer}.

In this paper, we will construct, by acting in a similar spirit, a model, in which by including into the action an appropriate interaction term, the universes represented by the massive bosonic particles moving in the minisuperspace (or according to the third quantization concept by the massive Klein-Gordon fields) can decay and collide with each other. Such decay processes will result in the emergence of the entanglement in the produced pairs of the universes.
We will also show that a model, with the abovementioned properties, can result from the non-minimally coupled biscalar gravity theory of varying fundamental constants in which both the speed of light and the gravitational constant are represented by the two non-minimally coupled scalar fields \cite{Balcerzak1}.

There have been many approaches to the idea of varying speed of light presented in the literature. Most of them, however, encounters profound conceptional problems such as violation of the Lorentz invariance \cite{Albrecht,Barrow1,Magueijo1,Clayton,Drummond,Clayton2}.
The theory of varying fundamental constants assumed in this paper is largely based on the locally Lorentz-invariant varying speed of light (VSL) theory  postulated by Magueijo in \cite{Magueijo1} that extends the VSL theories introduced in \cite{Albrecht,Barrow1} which break the general covariance  and, in consequence, require to choose a preferred reference frame (usually identified with the cosmological frame) to formulate the particular model. It was shown that such theories can solve the horizon, the flatness and the cosmological constant problem, however, the dynamics of varying speed of light is not given explicitly due to the lack of suitable terms in the action. The approach presented in \cite{Magueijo1} complements the mentioned VSL model by explicitly adding in the action the dynamical terms that govern the behaviour of the speed of light and the gravitational constant. It also proposes the definitions of covariance and local Lorenz invariance for the case of varying speed of light.
It was also shown that such a model includes a scenario in which the whole multiverse emerges from nothing \cite{Balcerzak2} and that the created pairs of the multiverses are entangled as confirmed by non-vanishing entanglement entropy \cite{Balcerzak3}. The interuniversal entanglement was also investigated in the context of the third quantized varying fundamental constants cyclic cosmological models in \cite{Robles_Bal} where it was argued that the third quantization naturally provides a thermodynamical description of the entanglement \cite{Alicki}. The emergence of the entanglement in pairs of created universes at the critical points of their evolution was investigated in \cite{Bellido}.

Our paper is organized as follows. In Sec. \ref{sec:1} we introduce the non-minimally coupled varying speed of light $c$ and varying gravitational constant $G$ cosmological toy model and define its third quantized action. In Sec. \ref{sec:2} we add to the action an interaction term that enables scattering and decay processes. In Sec. \ref{sec:3} we show that pairs of universes, which are produced in the decay processes induced by the interaction term introduced in Sec. \ref{sec:2}, are described by nearly maximally entangled Bell states. We also calculate the entropy of entanglement for such pairs. In Sec. \ref{sec:conc} we give
our conclusions.

\section{Third quantized non-minimally coupled varying $c$ and $G$ cosmological model}
\label{sec:1}

We start with a model of varying speed of light $c$ and varying gravitational constant $G$ introduced in \cite{Balcerzak2,Balcerzak1} where both fundamental constants in the original Einstein-Hilbert action are replaced with a certain functions of the two scalar fields. Thus, the resulting action formally describes a non-minimally coupled scalar-tensor gravity theory with two scalar degrees of freedom.
The considered model is largely based on the covariant and locally Lorentz-invariant varying speed of light theories postulated in \cite{Magueijo1} and is defined by the following action:
\begin{equation}
\label{action}
S=\int \sqrt{-g}  \left(\frac{e^{\phi}}{e^{\psi}}\right) \left[R+\Lambda + \omega (\partial_\mu \phi \partial^\mu \phi + \partial_\mu \psi \partial^\mu \psi)\right] d^4x,
\end{equation}
where $R$ is the Ricci scalar, $\Lambda$ is the cosmological constant, $\omega$ is the parameter of the model and $\phi$ and $\psi$ are the non-minimally coupled scalar fields, which values are, by definition, linked with the values of $c$ and $G$ via the following formulas:
\bea
c^3&=&e^{\phi}, \\
G&=&e^\psi.
\eea
By introducing the new fields $\beta$ and $\delta$ defined by:
\begin{eqnarray}
\label{fred}
\phi &=& \frac{\beta}{\sqrt{2\omega}}+\frac{1}{2} \ln \delta, \\
\psi &=& \frac{\beta}{\sqrt{2\omega}}-\frac{1}{2} \ln \delta,
\end{eqnarray}
the action (\ref{action}) can be recast into the Brans-Dicke type action which has the following form:
\bea
\label{actionBD}
S=\int \sqrt{-g}\left[ \delta (R+\Lambda) +\frac{\omega}{2}\frac{\partial_\mu \delta \partial^\mu \delta}{\delta} + \delta \partial_\mu \beta \partial^\mu \beta\right]d^4x. \nonumber \\
\eea
An introduction of varying $c$ into the action (\ref{action}) or (\ref{actionBD}) breaks the general covariance of the original theory \cite{Magueijo1}. This, on the other hand, forces one to choose a preferred reference frame in which our theory is formulated. We will follow an approach proposed in \cite{Magueijo1} and formulate our model in the cosmological frame defined by the flat FLRW metric which reads:
\be
\label{FLRW}
ds^2=-N^2(dx^0)^2+ a^2(dr^2 + r^2 d\Omega^2),
\ee
in which both the scale factor $a$ and the lapse function $N$ depend on the coordinate $x^0$. The form of the action (\ref{actionBD}) in the cosmological frame defined by the metric (\ref{FLRW}) reads:
\begin{eqnarray}
\nonumber
\label{action_s}
S &=& \frac{3 V}{8 \pi} \int dx^0 \left(-\frac{a^2}{N} a' \delta' - \frac{\delta}{N} a a'^2   + \Lambda \delta a^3 N  \right. \\
&-& \left.\frac{\omega}{2} \frac{a^3}{N} \frac{\delta'^2}{\delta}-\frac{a^3}{N}\delta \beta'^2 \right),
\end{eqnarray}
where $()'\equiv \frac{\partial}{\partial x^0}$.
Fixing the preferred reference frame completely requires to choose a specific form of the lapse function $N$. Since we are free in making such a choice, we will assume throughout the paper the following form of the lapse function $N$:
\be
\label{gauge}
N=a^3\delta.
\ee
The action (\ref{action_s}) can be further simplified by the application of the following sequence of the field transformations:
\be
X = \ln(a \sqrt{\delta}), \hspace {0.3cm} Y = \frac{1}{2A} \ln \delta
\ee
and
\be
\eta=r(AY-3X), \hspace {0.1cm} x_1=r(3Y-AX), \hspace {0.1cm} x_2=2\sqrt{\tilde{V}}\beta,
\ee
where $A=\frac{1}{\sqrt{1-2\omega}}$, $\tilde{V} = \frac{3 V}{8\pi}$ and $r=2\sqrt{\frac{\tilde{V}}{A^2-9}}$. The action (\ref{action_s}) in the new variables $\eta$, $x_1$ and $x_2$ takes the following form:
\be
\label{action_simple}
S= \int dx^0 \left[\frac{1}{4}(\eta'^2-x_1'^2-x_2'^2)+\bar{\Lambda}e^{-2\frac{\eta}{r}}\right],
\ee
where $\bar{\Lambda} = \tilde{V}\Lambda$.

The hamiltonian corresponding to the action (\ref{action_simple}) is:
\be
\label{ham}
H=\pi_\eta^2 -\pi_{x_1}^2- \pi_{x_2}^2-\bar{\Lambda}e^{-2\frac{\eta}{r}},
\ee
where $\pi_\eta=\frac{\eta'}{2}$, $\pi_{x_1}=-\frac{x_1'}{2}$ and $\pi_{x_2}=-\frac{x_2'}{2}$ are the respective conjugated momenta. Since both $\pi_{x_1}$ and $\pi_{x_2}$ are conserved quantities which is directly implied by (\ref{ham}) we can depict the classical evolution as a scattering of a point particle on the exponential potential
barrier.

Since we are interested in the high-curvature near singularity behaviour, which occurs for $\eta\rightarrow\infty$ (see Appendix \ref{app:A}), we need now to switch to the canonical quantum cosmological framework which is governed by the Wheeler-DeWitt equation. An application of the Jordan quantization rules which boil down to replacing the canonical momenta with the operators in accordance with the following scheme: $\pi_\eta\rightarrow \hat{\pi}_\eta=-i \frac{\partial}{\partial \eta}$, $\pi_{x_1}\rightarrow \hat{\pi}_{x_1}=-i \frac{\partial}{\partial x_1}$ and $\pi_{x_2}\rightarrow \hat{\pi}_{x_2}= -i \frac{\partial}{\partial x_2}$ leads to the following  the Wheeler-DeWitt equation:
\begin{equation}
\label{KG}
\ddot{\Phi} - \Delta \Phi + m^2(\eta) \Phi=0,
\end{equation}
where $\dot{( )}\equiv\frac{\partial}{\partial \eta}$, $\Delta = \frac{\partial^2}{\partial x_1^2}+\frac{\partial^2}{\partial x_2^2}$ and $m^2(\eta)= \bar{\Lambda} e^{-\frac{2}{r}\eta}$.

Eq. (\ref{KG}) is formally the same as the Klein-Gordon equation which allows us to perform the so-called third quantization of the Wheeler-DeWitt wave function $\Phi$. The third quantization procedure is completely analogous to the procedure of quantization of the Klein-Gordon field and leads to the Fock space whose vectors represent the states of the considered model of the multiverse. The first step of the third quantization procedure requires writing the so-called third quantized action which leads to the Wheeler-DeWitt equation. The third quantized action for the case of the Wheeler-DeWitt equation given by (\ref{KG}) has the following form:
\begin{eqnarray}
\label{3action}
S_{3Q}=\frac{1}{2}\int\left[ \dot{\Phi}^2 - (\nabla \Phi)^2 -m^2(\eta) \Phi^2 \right]d^2x d\eta,
\end{eqnarray}
where $\nabla$ is a two-dimensional gradient operator associated with the variables $x_1$ and $x_2$. The corresponding third quantized hamiltonian is:
\begin{eqnarray}
\label{3ham}
H_{3Q}=\frac{1}{2}\int\left[ \pi^2 + (\nabla \Phi)^2 + m^2(\eta) \Phi^2 \right]d^2x,
\end{eqnarray}
where the conjugated momentum $\pi=\dot{\Phi}$. The description of the classical evolution associated with the considered model can be find in the Appendix \ref{app:A}.

\section{Interacting universes in the minisuperspace}
\label{sec:2}

In our model, the set of universes is formally equivalent to the set of bosonic particles represented by the third quantized field $\Phi$, characterized by specific values of the momentum $p^\mu$. Since, in the quantum cosmological description, it is the minisuperspace that constitutes the true configurational space of the system under consideration, it seems natural to include an interaction term that allows for the decay and the scattering processes. The simplest action that enables the abovementioned processes reads:
\bea
\label{multi_act}
\bar{S}_M &=& \frac{1}{2}\sum_{i=1}^3\int \left\{\eta^{ab} \partial_a \Phi_i \partial_b \Phi_i - (m_i^2(\eta)) \Phi_i^2  \right\}d^2x d\eta \nonumber \\&-& g \int \Phi_1 \Phi_2 \Phi_3 d^2x d\eta,
\eea
where
\begin{equation*}
\eta^{ab} =
\begin{bmatrix}
1 & 0 & 0 \\
0 & -1 & 0 \\
0 & 0 & -1
\end{bmatrix},
\end{equation*}
$a,b=0,1,2$ and it enumerate the minisuperspace dimensions, $\partial_0\equiv \partial_\eta$, $\partial_1\equiv \partial_{x_1}$, $\partial_2\equiv \partial_{x_2}$ and $m_i(\eta)=\sqrt{\bar{\Lambda}}_i e^{-\frac{\eta}{r_i}}$ is the mass associated with the field $\Phi_i$. Since we are particularly interested in the description of the decay processes which are expected to occur in the high-curvature regime that takes place for sufficiently large value of the time parameter $\eta$ (see Appendix  \ref{app:A}), we will be using a simplified version of the action (\ref{multi_act}) which reads:
\bea
\label{multi_act_simp}
S_M &=& \frac{1}{2}\sum_{i=1}^3\int \left\{\eta^{ab} \partial_a \Phi_i \partial_b \Phi_i - m_{\Phi_i}^2 \Phi_i^2  \right\}d^2x d\eta \nonumber \\&-& g \int \Phi_1 \Phi_2 \Phi_3 d^2x d\eta,
\eea
with $m_{\Phi_i}=m_i(\eta_{h-c})$, where $\eta_{h-c}$ denotes the moment in which the high-curvature regime begins. Throughout the paper we will assume that $m_{\Phi_1}> m_{\Phi_2}+m_{\Phi_3}$.

The corresponding hamiltonian is:
\begin{eqnarray}
\label{multi_ham}
\hat{H}_{M}&=&\underbrace{\frac{1}{2}\sum_{i=1}^3\int\left[ \pi_i^2 + (\nabla \Phi_i)^2 +  m_{\Phi_i}^2 \Phi^2 \right]d^2x}_{\hat{H}_{free}} \nonumber \\&+& \underbrace{g\int\Phi_1 \Phi_2 \Phi_3 d^2x}_{\hat{H}_{int}},
\end{eqnarray}
where $\pi_i=\dot{\Phi}_i$ constitute a set of conjugated momenta.

\begin{figure}
\begin{center}
\resizebox{0.4\textwidth}{!}{\includegraphics{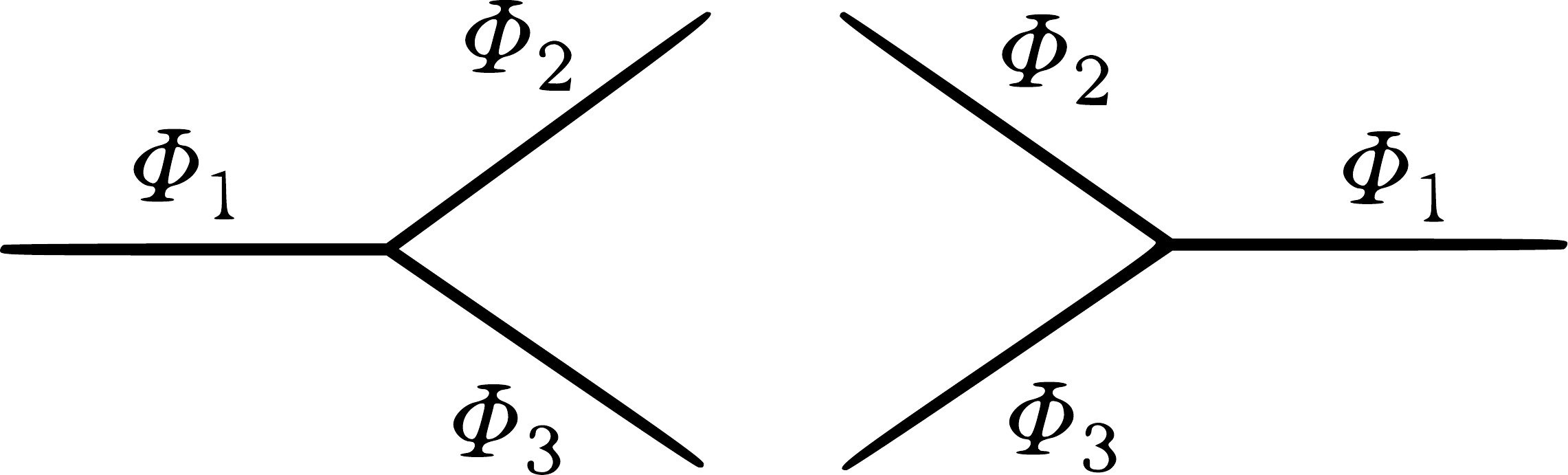}}
\caption{\label{diag1} Diagrammatic representation of the coupling (\ref{coupling}) up to order $g^2$.}
\end{center}
\end{figure}

\begin{figure}
\begin{center}
\resizebox{0.23\textwidth}{!}{\includegraphics{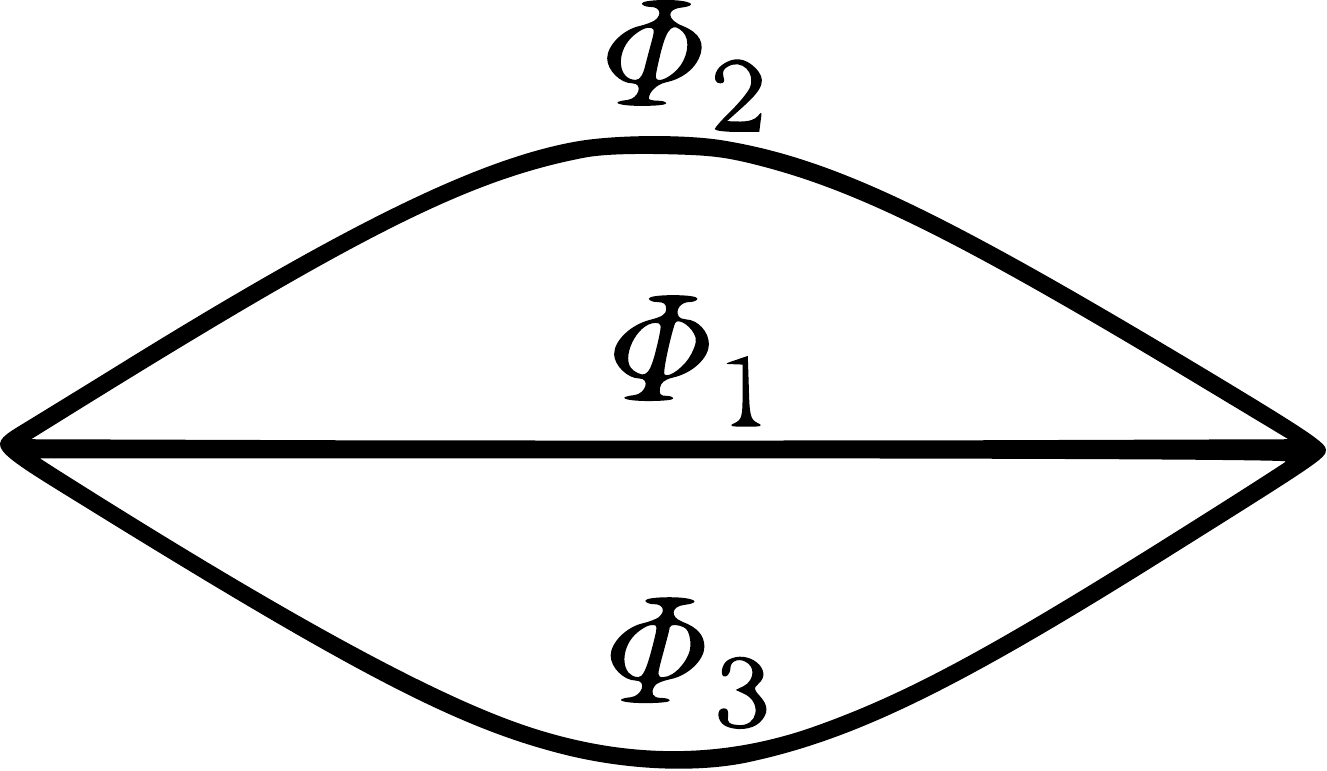}}
\caption{\label{diag2} Second order contribution to the vacuum energy.}
\end{center}
\end{figure}

The interaction term in (\ref{multi_ham}) couples the one-particle states of the field $\Phi_1$ with the definite energy and the definite momentum states of the two other fields, namely $\Phi_2$ and $\Phi_3$. Such coupling can more precisely be described by the following formula:
\be
\label{coupling}
\ket{1^{\Phi_1}_{\overrightarrow{0}}} \xlongleftrightarrow{\hat{H}_{int}} \ket{{\Phi_2}_{-\overrightarrow{p}}} \otimes \ket{{\Phi_3}_{\overrightarrow{p}}},
\ee
where $\ket{1^{\Phi_1}_{\overrightarrow{0}}}$ represents a zero-momentum one-particle state of the field $\Phi_1$ while $\ket{{\Phi_2}_{-\overrightarrow{p}}}$ and $\ket{{\Phi_3}_{\overrightarrow{p}}}$ represent the definite energy states (they are the eigenstates of $\hat{H}_{free}$) of the fields $\Phi_2$ and $\Phi_3$, respectively, with opposite momenta of magnitude $p$ (we assume the centre-of-mass reference frame). The presence of such interaction term allows for the processes in which a particle, represented by the field $\Phi_1$, decays into pairs of particles represented by the fields $\Phi_2$ and $\Phi_3$. The states $\ket{{\Phi_2}_{-\overrightarrow{p}}}$ and $\ket{{\Phi_3}_{\overrightarrow{p}}}$ are, on the other hand, coupled back to the state $\ket{1^{\Phi_1}_{\overrightarrow{0}}}$ (see Fig. \ref{diag1}).
The second order of the perturbations theory leads to the intermediate three-particle states (see Fig. \ref{diag2}) which contribute to the shift in the vacuum energy. Since this contribution influences only the phase factor which multiplies the one-particle state of the field $\Phi_1$ it will be neglected in the following considerations \cite{Holman,Bojanovsky}.

\section{Emergence of interuniversal entanglement via decay process. Bell states and the entropy of entanglement}
\label{sec:3}

It is convenient to use an interaction picture to describe the evolution of the state of the multiverse. Generally, the state of the multiverse $\ket{\Psi}$ can always be represented by linear combination of the eigenstates of the free hamiltonian $\hat{H}_{free}$:
\be
\label{state_lin}
\ket{\Psi}=\sum_n c_n(\eta) \ket{\varphi_n} e^{-iE_n \eta},
\ee
where $c(\eta)$ are some complex coefficients, $E_n$ and $\ket{\varphi_n}$ are the eigenvalues and the eigenstates of $\hat{H}_{free}$, respectively, and thus fulfil the following eigenequation:
\be
\label{H_f_eig}
\hat{H}_{free} \ket{\varphi_n} = E_n \ket{\varphi_n}.
\ee
By substituting Eq. (\ref{state_lin}) into the Schr\"{o}dinger equation
\be
\label{sch_full}
i \dot{\ket{\Psi}} = (\hat{H}_{free} + \hat{H}_{int}) \ket{\Psi}
\ee
one gets:
\be
\label{sch_full}
i \dot{c}_l= \sum_n c_n  \bra{\varphi_l} \hat{H}_I \ket{\varphi_n},
\ee
where $\hat{H}_I\equiv e^{i\hat{H}_{free} \eta} \hat{H}_{int} e^{-i\hat{H}_{free} \eta}$ is the interaction picture of $\hat{H}_{int}$.

For the case of the coupling given by (\ref{coupling})
we will be following the Wigner-Weisskopf approach \cite{Wigner} elegantly presented in \cite{Holman}. In order to make the notation more compact we will shorten the formula (\ref{coupling}) as follows:
\be
\label{coup_short}
\ket{G}\xlongleftrightarrow{\hat{H}_{int}} \ket{\alpha},
\ee
where $\ket{\alpha}$ constitutes the set of the eigenstates of $\hat{H}_{free}$ coupled to $\ket{G}$ via $\hat{H}_I$. We will also assume that the initial state of the considered setup is identical with  the state represented by $\ket{G}$, which is equivalent to the assumption that $c_G(0)=1$ and $c_{n\neq G}(0)=0$. The Wigner-Weisskopf approach for the case of coupling abbreviated by (\ref{coup_short}) gives the following expressions for the coefficients $c_G$ and $c_\alpha$:
\bea \label{evol_wigner1}
\dot{c}_G(\eta)&=&-\int_0^\eta d\eta' \Theta (\eta-\eta')c_G (\eta'),  \\ \label{evol_wigner2}
c_\alpha(\eta)&=& -i \int_0^\eta d\eta' \bra{\alpha}\hat{H}_I(0)\ket{G} e^{i(E_\alpha-E_G)\eta'} c_G(\eta'), \nonumber \\
\eea
where
\be \label{prop}
\Theta (\eta-\eta')=\sum_\alpha |\bra{G}\hat{H}_I(0)\ket{\alpha}|^2 e^{i(E_G-E_\alpha)(\eta-\eta')}.
\ee
It can be shown \cite{Bojanovsky} that:
\be \label{unitary}
|c_G(\eta)|^2 + \sum_\alpha |c_\alpha(\eta)|^2 = 1,
\ee
which means that the evolution is unitary even in the case of perturbative approach.

We will be considering the following scenario: as the evolution begins, the state of the universe is characterized by a high value of the curvature, since the value of the scale factor $a$ is close to zero. This, on the other hand,  corresponds to an infinite value of the time parameter, i.e. $\eta\rightarrow\infty$ (see Appendix  \ref{app:A}). In the state of high curvature, the interaction term $\hat{H}_{int}$ switches on, what enables the decay processes. We will argue that in such processes, the one-particle zero momentum state of mass $m_{\Phi_1}$  decays into a pair of particles of masses $m_{\Phi_2}$ and $m_{\Phi_3}$, maximally entangled in the momentum space.

The formula (\ref{prop}) adjusted to our particular setup is:
\bea \label{prop2}
\Theta (\eta-\eta')&=&\sum_{\vec{p}}|\bra{1^{\Phi_1}_{\overrightarrow{0}}}\hat{H}_I(0)\ket{{\Phi_2}_{-\overrightarrow{p}}} \otimes \ket{{\Phi_3}_{\overrightarrow{p}}}|^2 \times \nonumber \\ &\times&e^{i(m_{\Phi_1}-E_{\Phi_2}(p)-E_{\Phi_3}(p))(\eta-\eta')},
\eea
where $\vec{p}  \equiv \overrightarrow{p} $.
We will also define the following function:
\be \label{prop3}
D_0(\eta,\eta')\equiv\int_0^{\eta'}  d\eta'' \Theta (\eta-\eta''),
\ee
which derivative is:
\be \label{deriv}
\frac{\partial}{\partial \eta'}D_0(\eta,\eta')= \Theta (\eta-\eta').
\ee
By calculating the integral in Eq. (\ref{evol_wigner1}) by parts, one gets:
\bea \label{cphi}
&~&\frac{\partial}{\partial \eta} c_{\Phi_1}(\eta)= -\int_0^\eta d\eta' \Theta (\eta-\eta')c_{\Phi_1} (\eta')= \nonumber \\ &=& - D_0(\eta,\eta) c_{\Phi_1}(\eta) +\int_0^\eta d\eta' D_0(\eta,\eta') \frac{\partial}{\partial \eta'}c_{\Phi_1}(\eta'). \nonumber \\
\eea
The first term, in the formula above, is of second order in $\hat{H}_I$ while the second term, is of fourth order in $\hat{H}_I$. This means that up to the leading order the equation that governs the time evolution of $c_{\Phi_1}$ is:
\be \label{evol_master}
\dot{c}_{\Phi_1} = - D_0(\eta,\eta) c_{\Phi_1}.
\ee
The integral in Eq. (\ref{prop3}) which explicit form is given by:
\bea \label{de}
D_0(\eta,\eta')&=&\int_0^{\eta'}  d\eta''\sum_{\vec{p}}|\bra{1^{\Phi_1}_{\overrightarrow{0}}}\hat{H}_I(0)\ket{{\Phi_2}_{-\overrightarrow{p}}} \otimes \ket{{\Phi_3}_{\overrightarrow{p}}}|^2 \times \nonumber \\ &\times&e^{i(m_{\Phi_1}-E_{\Phi_2}(p)-E_{\Phi_3}(p))(\eta-\eta'')},
\eea
in the limit for $\eta \rightarrow \infty$ is:
\be \label{de1}
\bar{D}_0\equiv \lim_{\eta\rightarrow\infty}D_0(\eta,\eta)=i \Delta E_{\Phi_1}+\frac{1}{2} \Gamma,
\ee
where
\bea \label{phase}
\Delta E_{\Phi_1}&\equiv& P \sum_{\vec{p}}  \frac{|\bra{1^{\Phi_1}_{\overrightarrow{0}}}\hat{H}_I(0)\ket{{\Phi_2}_{-\overrightarrow{p}}} \otimes \ket{{\Phi_3}_{\overrightarrow{p}}}|^2}{m_{\Phi_1}-E_{\Phi_2}(p)-E_{\Phi_3}(p)}, \\
\Gamma&\equiv& 2\pi \sum_{\vec{p}} |\bra{1^{\Phi_1}_{\overrightarrow{0}}}\hat{H}_I(0)\ket{{\Phi_2}_{-\overrightarrow{p}}} \otimes \ket{{\Phi_3}_{\overrightarrow{p}}}|^2 \times \nonumber \\ &\times& \delta(m_{\Phi_1}-E_{\Phi_2}(p)-E_{\Phi_3}(p)),
\eea
$P$ denotes a principal value while $E_{\Phi_2}^2(p)=p^2+ m_{\Phi_2}^2$ and $E_{\Phi_3}^2(p)=p^2+ m_{\Phi_3}^2$. The solution of (\ref{evol_master}) in the limit $\eta \rightarrow\infty$ gives:
\be \label{master_sol}
c_{\Phi_1}= e^{-\bar{D}_0 \eta}=e^{-i \Delta E_{\Phi_1} \eta -\frac{1}{2} \Gamma \eta}.
\ee
In order to arrive to a more explicit version of the expression for $\Gamma$ we need first to calculate
\be
M(p)\equiv\bra{1^{\Phi_1}_{\overrightarrow{0}}}\hat{H}_I(0)\ket{{\Phi_2}_{-\overrightarrow{p}}} \otimes \ket{{\Phi_3}_{\overrightarrow{p}}}.
\ee
After decomposing all the three fields $\Phi_1$, $\Phi_2$ and $\Phi_3$ into modes with the following expression:
\bea \label{modes}
\Phi_i&=&\frac{1}{\sqrt{2\mathcal{A}}}\sum_{\vec{p}} \left(\frac{ a_{\vec{p}}}{\sqrt{E_{\Phi_i}(p)}}e^{-iE_{\Phi_i}(\vec{p})\eta} e^{i \vec{p} \cdot \vec{x}} \right. \nonumber \\
&+&
\left. \frac{a^\dagger_{\vec{p}}}{\sqrt{E_{\Phi_i}(p)}}e^{iE_{\Phi_i}(\vec{p})\eta} e^{-i \vec{p} \cdot \vec{x}} \right),
\eea
where $\mathcal{A}$ denotes the volume of compact  rectangular space-like part of the minisuperspace, the $M(p)$ assumes the following form:
\be \label{dyn}
M(p)=\frac{g}{\sqrt{8 \mathcal{A} m_{\Phi_1} E_{\Phi_2}(p) E_{\Phi_3}(p)}}.
\ee
In the three dimensional minisuperspace the explicit form of the expression for the $\Gamma$ is:
\be \label{gamma}
\Gamma=\frac{g^2}{8m^2_{\Phi_1}}.
\ee

The state of the three fields $\Phi_1$, $\Phi_2$ and $\Phi_3$ can be expressed as:
\bea \label{state}
\ket{\Psi(\eta)}&=& e^{-i \Delta E_{\Phi_1} \eta -\frac{1}{2} \Gamma \eta}\ket{1^{\Phi_1}_{\overrightarrow{0}};0_{\Phi_2};0_{\Phi_3}} \nonumber \\
&+&\sum_{\vec{p}}C_{\Phi_2\Phi_3}(p,\eta)\ket{0_{\Phi_1}} \ket{{\Phi_2}_{-\overrightarrow{p}}} \otimes \ket{{\Phi_3}_{\overrightarrow{p}}}.
\eea
The formula (\ref{evol_wigner2}) gives the following expression for the coefficients $C_{\Phi_2\Phi_3}(p,\eta)$ in (\ref{state}):
\be \label{coeff_2}
C_{\Phi_2\Phi_3}(p,\eta) = M(p) \frac{1-e^{-i\left(m_{\Phi_1,R}-E_{\Phi_2}(p)-E_{\Phi_3}(p)-i\frac{\Gamma}{2}\right)\eta}}{E_{\Phi_2}(p)+E_{\Phi_3}(p)-m_{\Phi_1,R}+i\frac{\Gamma}{2}},
\ee
where $m_{\Phi_1,R}=m_{\Phi_1}+\Delta E_{\Phi_1}$ is the renormalized  mass (in the following we skip the index $R$ in $m_{\Phi_1,R}$ and assume that $m_{\Phi_1}$ represents the renormalized mass $m_{\Phi_1,R}$). The expression (\ref{state}) explicitly describes the entanglement of the momentum eigenstates.

Since we are interested in calculating the entropy of entanglement we need first to calculate the reduced density  matrix associated with one of the universes. After tracing away the degrees of freedom of the universe represented by the field $\Phi_2$ we obtain the following reduced density matrix for the universe represented by the field $\Phi_3$:
\bea \label{dens_mat}
\rho_{\Phi_3}&=& \sum_{\vec{p}}\bra{{\Phi_2}_{\overrightarrow{p}}}\rho\ket{{\Phi_2}_{\overrightarrow{p}}}  = e^{-\Gamma \eta} \ket{1^{\Phi_1}_{\overrightarrow{0}}} \bra{1^{\Phi_1}_{\overrightarrow{0}}}\nonumber \\ &+&
\sum_{\vec{p}}|C_{\Phi_2\Phi_3}(p,\eta)|^2 \ket{{\Phi_3}_{\overrightarrow{p}}}  \bra{{\Phi_3}_{\overrightarrow{p}}},
\eea
where
$\rho\equiv\ket{\Psi(\eta)}\bra{\Psi(\eta)}$.

The von Neumann entropy is then given by:
\bea \label{entropyN}
\mathcal{S}(\eta)&=& - e^{-\Gamma \eta} \ln e^{-\Gamma \eta}
\nonumber \\ &-& \sum_{\vec{p}}|C_{\Phi_2\Phi_3}(p,\eta)|^2 \ln |C_{\Phi_2\Phi_3}(p,\eta)|^2.
\eea
In the narrow width limit $\Gamma << m_{\Phi_1},m_{\Phi_2}+m_{\Phi_3}$ the function $|C_{\Phi_2\Phi_3}(p,\eta)|^2$ is sharply picked at
\bea \label{momentum}
p_d&=&\frac{1}{2m_{\Phi_1}} [m^4_{\Phi_1}+m^4_{\Phi_2}+m^4_{\Phi_3}\nonumber \\
&-&2m^2_{\Phi_1}m^2_{\Phi_2}-2m^2_{\Phi_1}m^2_{\Phi_3}-2m^2_{\Phi_2}m^2_{\Phi_3 }]^{1/2},
\eea
so the last term in Eq. (\ref{entropyN}) can be approximated by the following expression:
\be \label{entropy_app}
\ln |C_{\Phi_2\Phi_3}(p_d,\eta)|^2 \frac{\mathcal{A}}{(2\pi)^2} \int dp^2 |C_{\Phi_2\Phi_3}(p,\eta)|^2.
\ee
In the narrow width limit we have that
\be \label{int_app}
\frac{\mathcal{A}}{(2\pi)^2} \int dp^2 |C_{\Phi_2\Phi_3}(p,\eta)|^2=1-e^{-\Gamma\eta},
\ee
what ensures fulfillment of the unitary evolution condition (\ref{unitary}). Since the average number of the universes with momentum of magnitude $p$ represented by the field $\Phi_2$ (or $\Phi_3$) is given by:
\begin{equation}
\label{TotNum}
\bra{\Psi(\eta)} a^\dagger_{\Phi_2}(p)a_{\Phi_2}(p) \ket{\Psi(\eta)}=|C_{\Phi_2\Phi_3}(p,\eta)|^2,
\end{equation}
the formula ($\ref{int_app}$) also gives the total number of the universes of either type ($\Phi_2$ or $\Phi_3$) produced in the volume $\mathcal{A}$. Thus, the decay results in a production of only one universe of either type in region $\mathcal{A}$.

\begin{figure}
\begin{center}
\resizebox{0.4\textwidth}{!}{\includegraphics{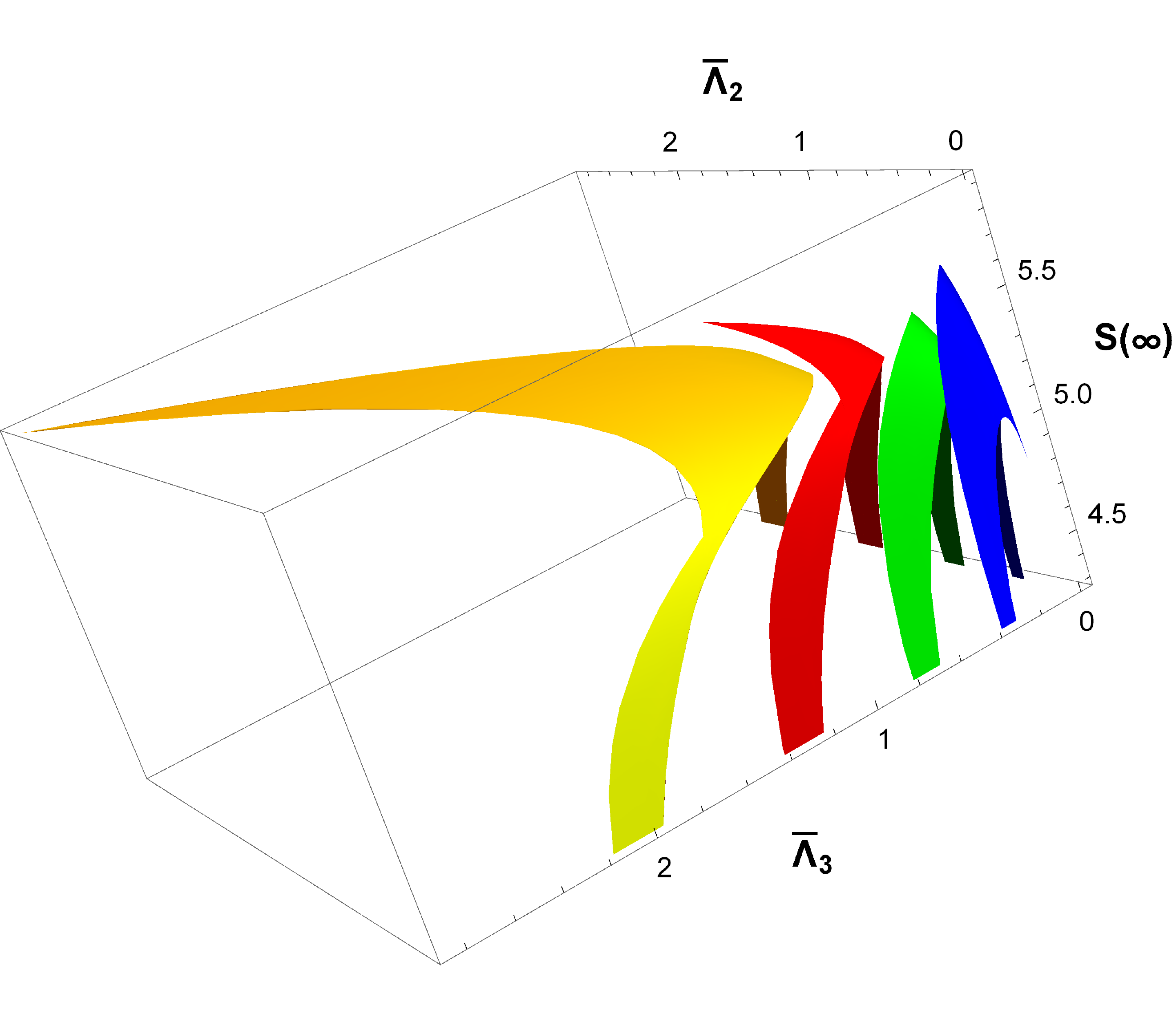}}
\caption{\label{entropy} Dependence of the entanglement entropy $\mathcal{S}(\infty)$ on the cosmological constants $\bar{\Lambda}_2$ and $\bar{\Lambda}_3$  of  pair of entangled universes produced in the decay processes. More elongated shapes corresponds to the higher value of the mass $m_{\Phi_1}$ of the parent universe.}
\end{center}
\end{figure}

Taking  into account the formula ($\ref{int_app}$) the von Neumann entropy (\ref{entropyN}) can be expressed as:
\be \label{entropy_2}
\mathcal{S}(\eta)= \Gamma \eta  e^{-\Gamma\eta}- (1-e^{-\Gamma\eta}) \ln |C_{\Phi_2\Phi_3}(p_d,\eta)|^2,
\ee
which in the high-curvature limit ($\eta \rightarrow \infty$, see Appendix \ref{app:A}) gives:
\be \label{entropy_2}
\mathcal{S}(\infty)= -\ln\left[ \frac{32 m^3_{\Phi_1}}{g^2 \mathcal{A} E_{\Phi_2}(p_d) E_{\Phi_3}(p_d)}\right].
\ee
The identification $E=i \frac{\partial}{\partial \eta}$ and the fact that asymptotically, for $\eta\rightarrow\infty$,  the value of $\pi_\eta = \sqrt{\bar{\Lambda}}$ (see Appendix  \ref{app:A}), allow  us to rewrite the formula (\ref{entropy_2}) as:
\be \label{entropy_3}
\mathcal{S}(\infty)= -\ln\left[ \frac{32 m^3_{\Phi_1}}{g^2 \mathcal{A} \sqrt{ \bar{\Lambda}_2 \bar{\Lambda}_3}}\right],
\ee
(see Fig. (\ref{entropy})).

Since in the considered case the function $|C_{\Phi_2\Phi_3}(p,\eta)|^2$ is narrowly peaked at $p=p_d$ a typical term that contributes to the expansion (\ref{state}) has the form:
\be
\label{Bell_state}
C_{\Phi_2\Phi_3}(p_d,\eta)\left[ \ket{{\Phi_2}_{-\overrightarrow{p}_d}} \otimes \ket{{\Phi_3}_{\overrightarrow{p}_d}}+\ket{{\Phi_2}_{\overrightarrow{p}_d}} \otimes \ket{{\Phi_3}_{-\overrightarrow{p}_d}}\right],
\ee
which means that the pair of the universes produced in the decay process is described by the nearly maximally entangled Bell state.

\section{Conclusions}
\label{sec:conc}
A trend in which the minisuperspace is treated as a fundamental scene, where the physical phenomena occur, gained a noticeable representation in the literature \cite{Bertolami,Pimentel,Marosek,Buonanno,Gasperini,R1,R2,R3,Barroso2}. Such an approach motivates an inclusion of different types of interactions between the Wheeler-DeWitt wave functions (representing individual universes of the multiverse) such as the interaction represented by the harmonic potential \cite{Serrano,Robles1,Kraemer}. Moreover, postulating such interactions and investigating the consequences of their existence seems to be the only way to make the concept of the multiverse a physical theory which is, at least in principle, amenable to observational verification.
Following this trend, we include in the action, stemming from the varying fundamental constants model, an interaction term that enables scattering and decay of the universes represented by the third quantized Wheeler-DeWitt wave function. The product of the decaying parent universe is a nearly maximally entangled pair of universes approximately described by a Bell state. Thus, the emergence of a nearly maximally entangled pairs of universes results from the inclusion of a very natural type of the interaction that allows for scattering and decay processes. Such type of entanglement is typical for the interaction term introduced in our multiverse toy model \cite{Holman}. However, the form of the state that describes the products of any decay process generically depends on the particular character of the interaction responsible for that decay \cite{Afik}. The entanglement concerns the spatial components of the minisuperspace momentum associated with individual universes. The strength of the entanglement measured by the von Neumann entropy depends on the values of the cosmological constants in each of the universes in the created pair and achieves higher values when both cosmological constants have similar values. Interestingly, the presented approach relies on the standard interpretation of the Fock space, which means that it treats the representation dependent orthonormal vectors, that form the basis in the Hilbert space of the multiverse, as vectors representing occupation with universes in a state completely determined by a particular set of proper quantum numbers. The standard interpretation of the Fock space was also assumed in \cite{Balcerzak2} where it was used to derive a scenario in which the whole multiverse subjected to the Bose-Einstein distribution emerged from nothing. On the other hand, there have been developed approaches to the problem of interuniversal entanglement, in which the standard quantum field theoretical interpretation of the Fock space was overridden in the sense that the Hilbert space basis vectors are assumed to define the excited states of the universes with a specific value of the momentum in the minisuperspace \cite{Balcerzak3,Robles_Bal}. Thus, the mechanism of interuniversal entanglement generation presented in this paper, seems to be quite a natural one since it does not assume any non-standard interpretation of the basic concepts of quantum field theory.

\appendix

\renewcommand{\theequation}{A.\arabic{equation}}
\renewcommand{\thefigure}{A\arabic{figure}}

\setcounter{equation}{0}
\setcounter{figure}{0}

\begin{appendices}

\section{Classical cosmological time evolution in the non-minimally coupled varying $c$ and $G$ model}
\label{app:A}
\begin{figure}
\begin{center}
\resizebox{0.4\textwidth}{!}{\includegraphics{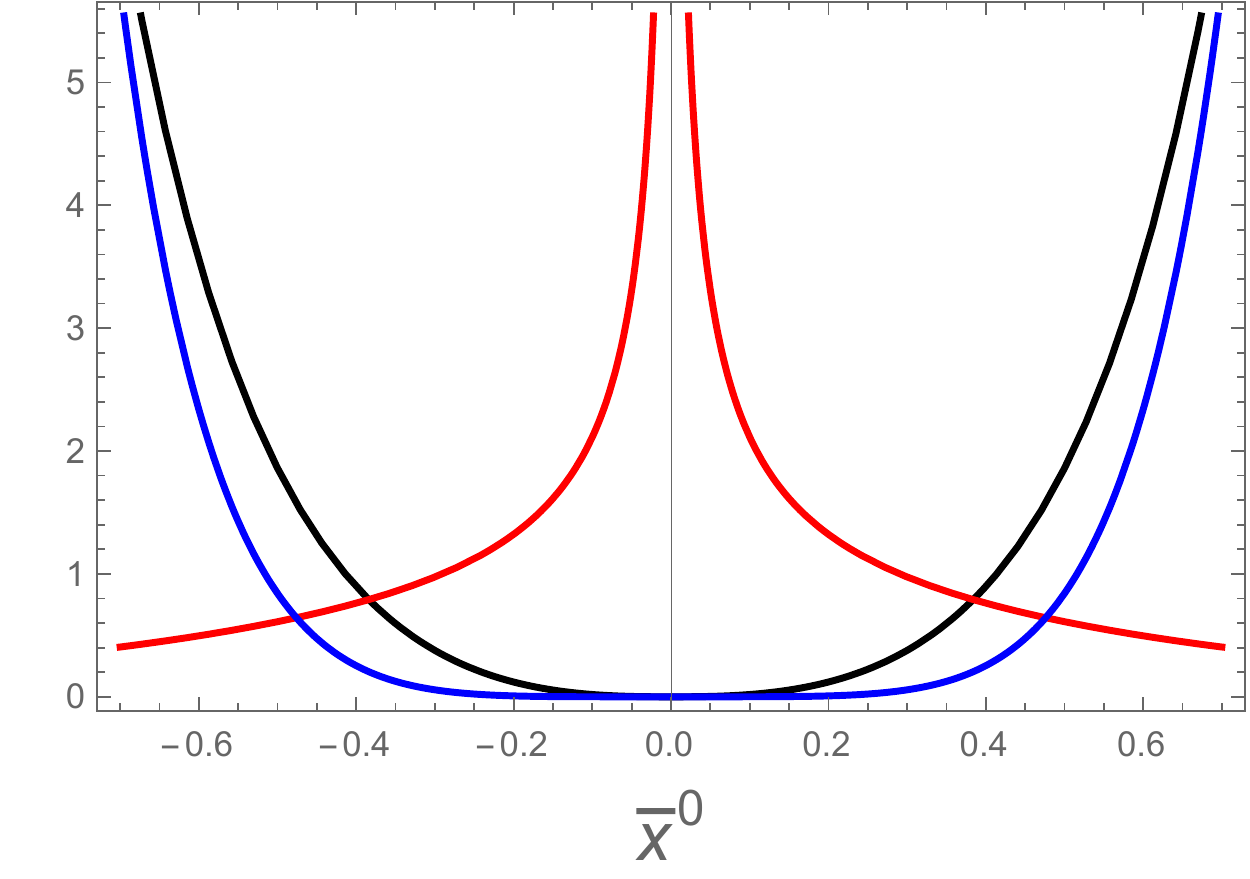}}
\caption{\label{acg} Time evolution of the scale factor $a$ (black), the speed of light $c$ (red) and the gravitational constant $G$ (blue) for $\bar{x}^0<0$  and  $\bar{x}^0>0$ which corresponds to the phase before and after the curvature singularity.}
\end{center}
\end{figure}

The classical evolution associated with the model defined in Sec. \ref{sec:1} is given by the following formulas \cite{Balcerzak1}:
\bea
\label{rozwio1}
a&=& \frac{1}{D^2 {(e^{ F x^0})}^2 \sinh ^ M |\sqrt{(A^2-9)\Lambda }x^0| },\\
\label{rozwio2}
\delta &=& \frac{D^6 {(e^{ F x^0})}^6}{\sinh ^ W |\sqrt{(A^2-9)\Lambda }x^0|},
\eea
where $M=\frac{3-A^2}{9-A^2}$, $W=\frac{2A^2}{9-A^2}$ while $D$ and $F$ are some integration constants. The variables $x^0$ is connected with the cosmological time $\bar{x}^0$ via the following expressions:
\begin{equation}
\label{conect}
\begin{split}
x^0 &= \frac{2}{\sqrt{(A^2-9)\Lambda}}  \arctanh \left(e^{\sqrt{(A^2-9)\Lambda}\bar{x}^0}\right)\,,
\hspace{0.2cm}
\text{for $\bar{x}^0<0$}\,,
\\
x^0 &= \frac{2}{\sqrt{(A^2-9)\Lambda}}  \arctanh \left(e^{- \sqrt{(A^2-9)\Lambda}\bar{x}^0}\right)\,,
\hspace{0.2cm}
\text{for $\bar{x}^0>0$}\,,
\end{split}
\end{equation}
where we assumed that $A^2>9$. The set of the solutions above defines the time evolution of the scale factor $a$, speed of light $c$ and the gravitational constant $G$. The qualitative behaviour of the three parameters is depicted in Fig. (\ref{acg}). It can be seen that the model contains the pre-big-bang contraction (for  $\bar{x}^0<0$) and the post-big-bang expansion (for  $\bar{x}^0>0$) with both phases separated by the curvature singularity at $\bar{x}^0=0$.

The Hamilton's equation of motion associated with the hamiltonian (\ref{ham}) gives the following time evolution:
\bea
\label{ham_sol1}
\eta&=&r \ln \sinh|\sqrt{(A^2-9)\Lambda }x^0|, \\
\label{ham_sol2}
x_1&=& -2 \pi_{x_1} x^0 + E, \\
\label{ham_sol3}
x_2&=&-2 \pi_{x_2} x^0 + P,
\eea
where $E$ and $P$ are some integration constants. The solution (\ref{ham_sol1}) allows us to define the high-curvature limit for $\eta\rightarrow \infty$ which corresponds to near curvature singularity evolution and the low-curvature limit for $\eta\rightarrow -\infty$ which corresponds to the late evolution (far form the curvature singularity). The asymptotic value of the momentum $\pi_\eta$ in the high-curvature limit (for $\eta\rightarrow \infty$) is:
\be
\label{moment_as}
\pi_\eta = \sqrt{\bar{\Lambda}}.
\ee

\end{appendices}

\end{document}